\documentclass[times,twocolumn]{IEEEtran} 
\usepackage{color,times,fullpage} 
\usepackage{epsfig,url,amsmath}
\usepackage{setspace}
\usepackage{latexsym}
\usepackage{graphicx}
\usepackage{times}
\usepackage[bf]{subfigure}

\newcommand{\rednote}[1]{\textcolor{red}{ \em #1 }}
\newcommand{\bluenote}[1]{\textcolor{blue}{ \em #1 }}
\renewcommand{\rednote}[1]{}
\renewcommand{\bluenote}[1]{}

\begin{document}
\title { TARMAC: Traffic-Analysis Reslient MAC Protocol for Multi-Hop Wireless Networks }

\author {Ke Liu,  Adnan Majeed  and Nael B. Abu-Ghazaleh \\ 
CS Dept., SUNY, Binghamton NY 13902 \\ 
\{kliu,adnan,nael\}@cs.binghamton.edu }
\date{}

\maketitle
\thispagestyle{empty}

\begin{abstract}
Traffic analysis in Multi-hop Wireless Networks can expose the
structure of the network allowing attackers to focus their efforts on
critical nodes.  For example, jamming the only data sink in a sensor
network can cripple the network.  We propose a new communication
protocol that is part of the MAC layer, but resides conceptually
between the routing layer and MAC, that is resilient to traffic
analysis.  Each node broadcasts the data that it has to transmit
according to a fixed transmission schedule that is independent of the
traffic being generated, making the network immune to time correlation
analysis. The transmission pattern is identical, with the exception of
a possible time shift, at all nodes, removing spatial correlation of
transmissions to network strucutre. Data for all neighbors resides in
the same encrypted packet.  Each neighbor then decides which subset of
the data in a packet to forward onwards using a routing protocol whose
details are orthogonal to the proposed scheme.  We analyze the basic
scheme, exploring the tradeoffs in terms of frequency of transmission
and packet size.  We also explore adaptive and time changing patterns
and analyze their performance under a number of representative
scenarios.
\end{abstract}

\section{Introduction}

Technological advances in VLSI, MEMS, and wireless communication have
ushered in a new age of miniature, low cost, low-energy,
micro-sensors.  Networks of such devices, called Wireless Sensor
Networks (WSNs), hold the promise of revolutionizing sensing across a
range of civil, scientific, military and industrial applications. This
emerging technology provides an opportunity to collect information at
unprecedented resolution due to their low cost, small size, ease of
deployment, and ability to provide fine-grained up-close sensing. It
is anticipated that instrumenting physical environment with thousands
of tiny sensing devices can significantly improve our understanding of
the real world and provide one of the necessary interfaces between the
physical and digital worlds.

Traffic analysis can be of major concern to secure multi-hop wireless
networks.  Since packets are transmitted over open air, they can be
easily intercepted.  The use of encryption protects the
confidentiality of the data.  However, the network becomes vulnerable
to traffic analysis where an adversary can extract the pattern of
communication if not its contents.  Knowing the pattern of
communication can be used to guide more effective attacks on the
network.  For example, the attacker can jam a base station or critical
relay nodes to cripple large parts of the network with a modest
effort.  Further, the information about what nodes are communicating
can itself be sensitive.  For example, in a military scenario, if an
adversary observes increased sensor activity in an area, it may deduce
that its assets there have been discovered and move them before they
are attacked.  

The main enabler for traffic analysis is that data communication is
event driven.  Thus, when a node has data to send, it sends it
immediately, subject to its MAC layer protocol.  An attacker, can
derive the presence of data from the time correlation of packet sends.
Further, it can extract the spatial communication pattern by tracking
which relay nodes forward a packet onwards.

Thus, to protect from
traffic analysis, the data traffic must be engineered to make it
impossible to extract temporal or spatial correlation of physical
transmission to the presence of data.  We discuss some of the existing
related work in Section~\ref{related}.

The primary contribution of this paper is a Traffic-Analysis Resilient
MAC protocol (TARMAC).  In TARMAC the traffic generated by
each node is independent of data (and control packets) to be sent by
it.  Instead each node sends encrypted frames, each consisting of
multiple packets, in a pattern that is the same across all nodes, with
the exception of a possible time shift that is independent of the
data.  The generated frame is broadcast locally without a specific
destination and is received by all neighbors.  The size of the frame
is independent of the amount of actual data it contains (for example,
it may be constant in size).  Each frame consists of all the data the
node has to transmit to all its neighbors (up to the size of the
frame).

Neighboring nodes pick up the broadcast frame, and may decide to
forward portions of it onwards when it is their time to send according
to the routing protocol (which is orthogonal to this functionality,
but as we show later, could gain from being aware of the transmission
pattern).  Thus, each packet may contain multiple pieces of data
destined to different locations.  Each receiving node may forward a
subset of the data packets in a frame, combined with other packets
from other nodes, in the next frame it generates.  

A useful analogy is a vehicle transportation system.  Existing MAC
protocol have data packets each using their own car (even-driven
transmission), allowing an intruder to simply observe the presence of
a car to detect data transmissions and to follow the car's progression
to extract spatial pattern of communication.  In TARMAC, there are
only buses that leave to preset locations at scheduled intervals.  At
every stop, passengers from different arriving buses may hop onto
other buses that take them towards their destination.  To an outside
observer, the buses always leave on schedule, regardless of what
passengers are inside them (or in fact, whether there are any
passengers at all). %% Further, by making the bus schedule uniform at

Different routing protocols and, in fact, link layer ones can
inter-operate with TARMAC.  The routing protocol defines how the data
packets are processed at the next hop (how passengers move from bus to
bus).  For example, the source may indicate who the next hop is, say
by geographic routing, and in that case, only the designated next hop
forwards a piece of data.  Any other routing protocol can be used
instead.  Similarly, Link level reliability may be implemented by
having the next hop acknowledge a packet when it is its time to
broadcast. Alternatively, as each node broadcasts packets the next hop
forwarding a packet can act as an implicit acknowledgement.  The
scheme is described more formally in Section~\ref{bus}.

Clearly, analysis resilience comes at a cost.  For example, having to
follow pre-set schedules implies that average packet delay will be
higher; passengers that use public transportation expect a longer
travel time than those with their own car.  Further, to make the
structure of the network resistant to analysis, the pattern of
transmission should be independent of the location in the network;
otherwise, the pattern is open for spatial analysis where critical
nodes are identified by their level of activity.  In contrast to
traditional bus systems, where buses merge at intermediate hubs that
have much higher activity than edge stops, for passenger exchange,
this model requires over-provisioning such that even distant stops act
as hubs.  As a result, it is likely that many of the frames near these
edge stops do not have many passengers (otherwise, hubs wont have
enough buses to carry their passengers).  We analyze these tradeoffs
and explore the effect of critical parameters such as send rate, frame
size and traffic pattern on the performance of the basic TARMAC.

We also consider approaches for improving the efficiency of the scheme
in Section~\ref{improve}.  More specifically, we explore adapting
the transmission pattern (uniformly) to match the desired capacity.
Further, we explore time-varying transmission schedules where only
portions of the network display high transmission rates at a given
time.  The portions of high activity change over time such that all
nodes still have the same transmission pattern (only shifted in time).
The routing protocol is aware of the changing capacity and can direct
packets to areas of the network that currently have high capacity.  In
Section~\ref{experiments} we evaluate the basic scheme, as well as these
improvements to it.  Finally, Section~\ref{conclude} presents some
concluding remarks.

\section{Background and Related Work}\label{related}

\rednote{Adnan, you should not use references as an object/part of the sentence.  You can say ``Smith et al~\\cite\{smith\}'' or ``Intrusion detection in the context of MHWNs~\\cite\{smith\} is a ...''}

\rednote{Can refer to classical traffic analysis in general networks--
onion routing etc..}

Deng et al ~\cite{deng-intrusion} discuss the traffic analysis problem
as it relates to Multi-Hop Wireless Networks (MHWNs).  They isolate
the following properties that enable traffic analysis: (1) Time
correlation between receiving and forwarding a packet.  Using this
correlation an adversary can track the progression of a packet and
extract the structure of the connections and network; (2) Unencrypted
packets: if packets are plain text, they can be studied to find the
next hop and this can be done successively to get to the base-station;
and (3) Areas of interest and areas near the base-station generally
have higher traffic, the activity level can then be used to identify
such key points.  Resilience to traffic analysis requires eliminating
all three of the above ingredients.

Traffic analysis has been studied in the context of wired and wireless
networks (e.g.,~\cite{mathewson-practical, newman-metrics}). In wired
networks onion routing~\cite{syverson97anonymous} is the prominent
anti-traffic analysis protocol. It camouflages the destination by
routing through intermediate proxies. In wired networks, however,
hop-to-hop behavior is not in question and only the source and
destination are being protected. In warless networks, the channel is
broadcast in nature and there is a correlation between receiving a
packet and sending it and this can be used to determining flow
directions. Also, since wireless networks generally have low duty
cycles sudden increase in sending rates at a particular location would
mean an event has taken place at that location.

Deng et al propose the use of multi-path multi-base-station routing to
protect against traffic analysis ~\cite{deng-intrusion}.  Data packets
flow up routing trees towards the base station. To prevent traffic
analysis each node transmits at fixed time intervals irrespective of
the data present. If a node has data, it forwards it in it alloted
time slot, otherwise it just forwards dummy packets.  If a node does
not hear its packet forwarded, it keeps sending the same packet when
its time slot arrives. This scheme is different from TARMAC in a
number of ways.  TARMAC takes advantage of the broadcast nature of the
wireless channel to group all outgoing packets from a node together.
Deng's approach attempts to reduce hot-spots by limiting flows feeding
into a hot-spot.  They do not protect control packets that need to be
flooded in the network.  The latency is much higher for this approach
because a node transmits a packet to its parent only when it starts
transmitting dummy packets or it over hears the parent node
transmitting its packet.  Every node has to change the encryption so
that packets are not followed easily by an adversary. In TARMAC, the
packets are disassembled and reassembled at each node; thus, it does
not require that a unique key be applied at every hop.  

In a followup work~\cite{deng-antitrafficanalysis} the authors propose
randomizing the path taken by every packet so that a pattern is not
found. They also proposes having some packets in addition to their
path take random fake paths to take an adversary in the wrong
direction. The paper goes on to propose generating random areas of
high traffic (hot-spots) to misguide an adversary.  This solution is a
routing level solution with different properties than TARMAC.  To our
knowledge these are the only works on traffic analysis in sensor
networks.  We compare TARMAC against both of these works in Section~\ref{eval}.

\section{TARMAC: Basic Scheme}\label{bus}

\rednote{Good start.  Recommend:
-comparative analysis to the existing schemes
-identification of the tradeoffs present in the design (size of
packet, time-period, etc..).  Present the discussion of the drawbacks
in a way that leads into the next section (the improvements), such
that the next section can start with something like: ``The basic scheme
has two major drawbacks, inability to adapt to the level of traffic
and excessively high overhead.  In this section, we address these
drawbacks by...''}

\rednote{To make description easier and more accurate, maybe we should
use packet for data, and frame for the buses}

In this section, we present the basic TARMAC scheme.  We also discuss
tradeoffs and expected behavior of this base model.  The drawbacks of
the model in terms of energy efficiency and increased contention pave the way for the suggested improvements in the next Section.

\subsection{Basic TARMAC Scheme}

TARMAC is a MAC layer protocol that takes in to consideration the
makes traffic resistant to analysis by using uniform (but possibly
time-shifted) transmission schedules at all nodes. TARMAC emulates the
communication network alternative of a city bus system.  TARMAC uses
encrypted frames that are sent at pre-scheduled times (it can
inter operate with different link/MAC layers), irrespective of the
presence of data packets.  Different outgoing data packets, possibly
targeted to different destinations and next hops, are placed in the
same TARMAC frame; any empty slots are left empty.  Note that empty
slots are hidden by the encryption and the occupancy of each frame is
not visible to the attacker.  Each receiver of the frame can examine
it and decide which packets are its responsibility to forward:
virtually any routing protocol can be used to establish routing
responsibilities.

  TARMAC frames resemble public buses in that they have a fixed number
of slots and leave at pre-scheduled times, with passengers
(data-packets) taking up seats in the bus.  In typical public
transportation systems, provisioning is not uniform: for example,
stops at remote routes have much lower bus activity and perhaps
smaller buses than a central hub where many routes converge.  Thus,
extracting the structure of the public transportation network by
simply observing the level of activity at a stop becomes possible.

To protect against such analysis TARMAC requires that all nodes follow
the same schedule (later, we consider the possibility of a
time-shift).  {\em The basic TARMAC scheme we discuss in this section
is one where all the nodes transmit with a fixed period and fixed size
packets}; later we explore relaxing this model.  Having the
transmission times and sizes be uncorrelated to the data transport
being carried through TARMAC makes it impossible to detect the
presence of, for example, event based data.  Furthermore, the
broadcast nature of the wireless medium hides information about the
receiver of the packet; thus, a TARMAC frame is like a number of
concurrent buses that leave to each of the one-hop neighbors.

In contrast to these advantages of TARMAC, conventional MAC transmit
data only when there is data to transmit, providing valuable
information to attackers.  Further, a data packet can be followed as
it is retransmitted by intermediate hops to extract the full
connection and the eventual destination.

\subsection{Basic Parameters and Provisioning}

In the basic model with periodic equal size frame, the relevant
parameters are the transmission period $\tau$ and the size of the
frame in slots $s$.  The capacity, measured in slots per node per
second, for the basic TARMAC can theoretically be expressed as
$\frac{s}{\tau}$.  Within the physical limitations of the channel,
increasing either $s$ or reducing $\tau$ leads to increasing capacity
by either sending larger packets or sending packets more often
respectively.  However, sending smaller frames more frequently fosters
shorter delays, but increases frame overhead.  In addition, larger
size frames are more vulnerable to collisions and transmission losses.

A tension between the capacity of the network and the energy
efficiency of TARMAC arises.  At one extreme, all nodes may be made to
appear like a remote station, leading to energy efficiency but loss of
capacity since the bottleneck nodes now do not have sufficient
capacity to carry the offered load.  On the other extreme, the network
may be provisioned so that all nodes are transmitting at a sufficient
rate to enable the bottleneck nodes to continue to forward the
traffic.  This leads to excessive overhead in remote or idle areas.
These and other drawbacks are discussed in the next subsection.

\subsection{Drawbacks of Basic TARMAC}

To be able to carry the required traffic, the basic TARMAC must be
provisioned sufficiently such that $\frac{s}{\tau}$ is greater or
equal the required traffic at bottleneck nodes.  However, provisioning
for the worst case has the following drawbacks:
\begin{itemize}
\item {\em Over or under-provisioning:} it is difficult to predict
what the maximum bottleneck capacity is for some networks.  The
maximum reporting rate may be difficult to predict or the deployment
may be ad hoc making a-priori analysis of bottleneck nodes difficult.
As a result, the choice of $s$ and $\tau$ may lead to insufficient
capacity to carry the reported data and leading to increased delay and
loss of data as buffer sizes grow.  Alternatively, it may lead to more
aggressive sending and loss of efficiency as most slots remain idle.

\item {\em Possible High Energy Cost:} By provisioning to the rate of
the expected bottleneck, most of the nodes will be transmitting at a
rate higher than that needed to carry their traffic.  We note here
that as long as an average occupancy of more than one slot per frame
the total number of transmissions will be reduced.  However, the size
of each frame will be likely be bigger than the size of the individual
data packets due to over-provisioning; some savings in framing and
protocol overhead may result from combining multiple packets into a
single transmission.

\item {\em Low maximum throughput:} As the node capacity is increased
by reducing the period or increasing the size, all nodes start sending
more aggressively and the contention level increases.  This includes
nodes that do not have high occupancy, limiting the maximum throughput
that can be achieved by nodes that do have data to send.

\item {\em Increased delay:} Since each node is not forwarding the
packets as soon as it gets it, there may be a larger packet delivery
latency as packets wait for the next frame transmission at every
intermediate hop.
\end{itemize}

\noindent
In the next section, we identify a number of improvements to the base
TARMAC that address some of these drawbacks.

\section{Improvements and Extensions}\label{improve}

In this section we discuss a number of improvements to the basic
TARMAC in response to the drawbacks identified in the previous section.  
\subsection{Adaptive TARMAC}

Static provisioning of TARMAC in terms of $s$ and $\tau$ can lead to a
TARMAC configuration that is either insufficient to sustain the
reported data, exceeds even the bottleneck requirements, or
unsustainable in some areas of the network due to exceeding the
capacity of the channel.  We propose an adaptive version of TARMAC
that allows adjustment of the reporting rate or bus size and works as
follows.  
\begin{itemize}
\item Increasing capacity: When a node detects that its traffic
demands exceed its capacity, it requests an increase in the reporting
rate.  The increase is requested from the base-station (or a rate
regulation leader), which can periodically adjust the overall rate
based on the observed behavior.  Adjustment is initiated through a
flood packet (within the TARMAC mechanism) that proposes a new rate
and a time when the nodes should switch to it.  Precise
synchronization is not needed.  The nodes then all switch to the new
rate.

\item Reducing capacity: Capacity should be reduced in two cases: (1)
a TARMAC rate exceeds local channel capacity; and (2) the TARMAC rate
is too high and exceeds the need of all nodes.  In the first case,
nodes may elect to reduce capacity if the current TARMAC rate locally
exceeds the available bandwidth.  This will occur in high density
areas.  A reduction in schedule in such a condition is allowed because
it does not reflect information about the traffic, but only the local
density.  In the second case, several options are possible, including
tracking the maximum occupancy observed by packets along the way, and
informing the regulator if the maximum of these is low.
\end{itemize}
\noindent
Note that adapting the transmission activity to the bottleneck load
allows attackers to detect the bottleneck level of activity of the
network (but not where this bottleneck is).  However, we believe that
the advantages in terms of capacity and efficiency makes exposing this
information worthwhile.

\subsection{Use of Multi-Path Routing}

High energy expenditure and low effective capacity may arise in basic
TARMAC due to frames that are sparsely populated, wasting energy and
occupying available channel time, leaving less of it available for
frames carrying data.  In response to this problem, we propose that
the routing protocol should be aware of the TARMAC characteristics.
For basic TARMAC, this entails the use of multi-path
routing~\cite{lee-01} to take advantage of the available capacity
in nearby nodes.  In conventional multi-hop wireless networks, the use
of multi-path routing often does not lead to appreciable improvement
in capacity because some of the links making up the different paths
may be in interference range with each other, reducing the available
bandwidth for each.  For TARMAC, this is not the case because
multi-path routing simply takes advantage of the available slots and
does not require any additional transmissions.

\subsection{Non-uniform TARMAC Schedule}

Thusfar, we have assumed that the identical schedule maintained by all
the nodes is a simple periodic one where all nodes transmit at the
same period and with the same size.  As we have seen, these leads to
uncontrolled contention, especially in areas that are dense.  However,
an interesting possibility is to allow periodic schedules that do not
have a fixed transmission rate.  For example, nodes may alternative
between a high send rate and a low send rate state.  At a given
instance, only some of the nodes are in the high rate portion of their
schedule and contention is reduced.  More complex schedules, as well
as the possibility of adapting the schedule, are also possible.  All
nodes have the same schedule, with the exception of a time-shift,
retaining the desirable traffic-analysis resilience features.

Within these general parameters a large solution space emerges based
on the type of the schedule, and coordinating which nodes are in the
high send rate together.  These sets of nodes must be sufficient to
provide connectivity and capacity to carry the desired traffic.
Further, since the set of high rate nodes (which are most suitable as
relay nodes) changes with time, the routing protocol should be able to
adaptively use the high rate nodes.  

A promising example of this class of TARMAC is one where nodes
self-organize into clusters and synchronize activity periods such that
at least one node is active at any given time.  Routing based on zone
addresses (e.g., as in GAF~\cite{xu-01}) is then used to
automatically use the current high rate node as the forwarding node
for connections going through these area.  As nodes change schedules
they change forwarding responsibilities; nodes about to become
forwarders should retain recent packets that have not been forwarded
yet so that they can forward them when they switch to the high rate
mode.  We do not study this flavor of TARMAC in this paper.  We intend
to do so in a follow-up paper in the near future.

\section{Experimental Evaluation}\label{eval}
\label{experiments}
\rednote{Can you write an outline of what experiments/data you are
planning to represent?  Experimental planning is a critical part of a
good paper.}

In this section, we present an experimental evaluation exploring the
performance of TARMAC under different conditions, and compare it
against two existing schemes for protection against traffic analysis
in sensor networks\cite{deng-intrusion,deng-antitrafficanalysis}; we
call these solutions Intrusion 1 and Intrustion 2 respectively.  We
also compare against an unprotected network to evaluate the overhead
necessary for camouflaging traffic.  We implemented TARMAC in network
simulator NS2 (version 2.29) \cite{ns2}, by extending IEEE 802.11 MAC
protocol according to our design.  To enable fair comparison, we also
implemented Intrusion 1 and 2 on NS2 (they were implemented on TinyOS
originally).

In our experiments, 100 sensors are regularly deployed in $10\times10$ grids covering an
area of $200\times200 meter^2$, in which each node is located at the center of each grid.
The table \ref{tab:simpara} shows some simulation parameters general for all studies.
Other simulation parameters are summarized in Table~\ref{tab:simpara} unless
explicitly stated.

\begin{table}
\centering
\begin{tabular}{|c|c|}
\hline Parameter & value \\
\hline No. of Data Sources & 100 \\
\hline Traffic type & CBR\\
\hline CBR packet size & 32B\\
\hline Transmission Range & 40 meters\\
\hline BandWidth & 2Mbps\\
\hline Routing Period & 5 seconds\\
\hline Traffic Period & 100 seconds \\
\hline Simulation Period & 400 seconds \\
\hline TARMAC slot size & 64 Byte\\
\hline
\end{tabular}
\caption{Some simulation parameters}
\label{tab:simpara}
\end{table}

\subsection{Analysis of Basic TARMAC}

\begin{figure}[h]
\centering
\includegraphics[width=0.45\textwidth]{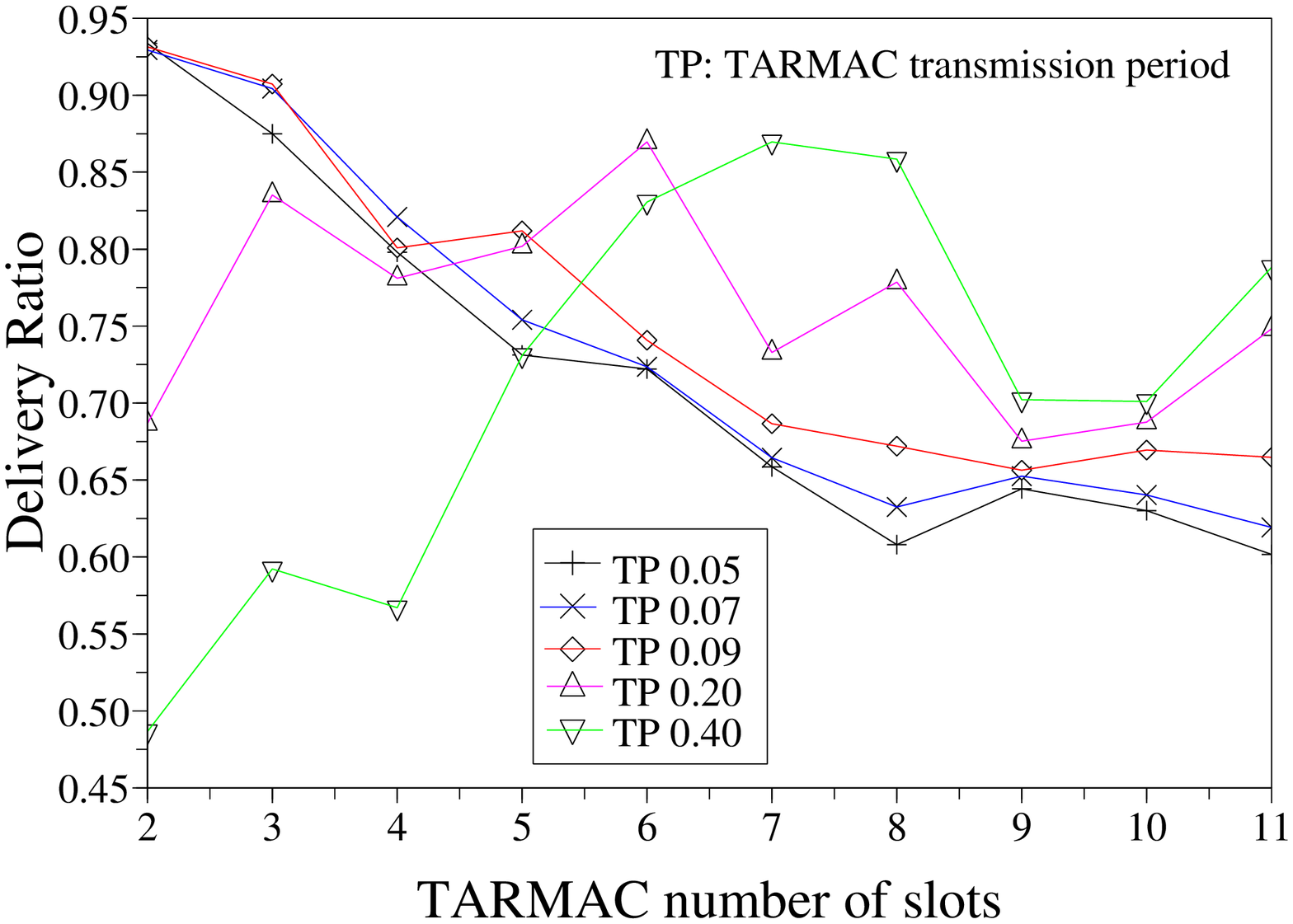}
\caption{TARMAC Delivery Ratio Study}
\label{fig:busslots-dr}
\end{figure}
\begin{figure}[h]
\centering
\includegraphics[width=0.45\textwidth]{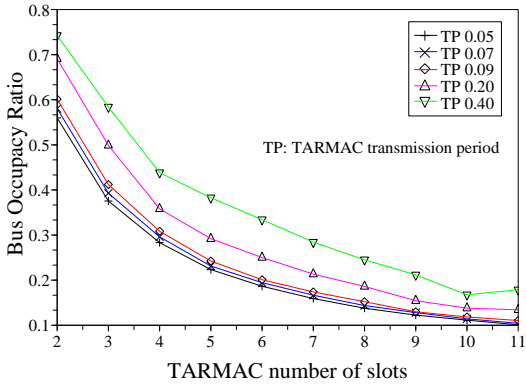}
\caption{TARMAC Bus Occupation Study}
\label{fig:busslots-occupacy}
\end{figure}
 \begin{figure}[h]
 \centering
 \includegraphics[width=0.45\textwidth]{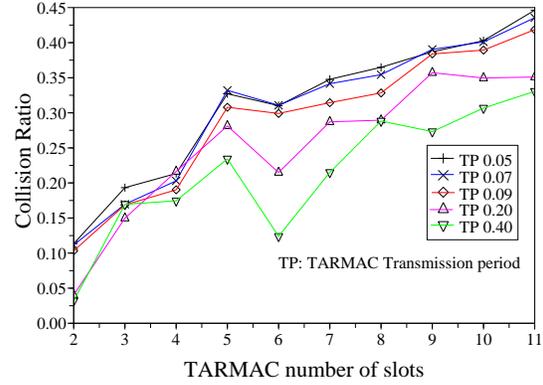}
 \caption{TARMAC Collision Ratio Study}
 \label{fig:busslots-coll}
 \end{figure}

Figure \ref{fig:busslots-dr} shows the delivery ratio of shortest path
routing on TARMAC as a function of the size of the frame and the
transmission period. When TARMAC has a small period, even when the
buses are small, the delivery ratio is high since the collision ratio
is low as bigger frames suffer more collisions
(Figure~\ref{fig:busslots-coll}).  When the period is large, if the bus
size is small, then the effective capacity is not sufficient to carry
the packets leading to increased packet delay and reduced delivery
ratio.

\begin{figure}[h]
\centering \includegraphics[width=0.45\textwidth]{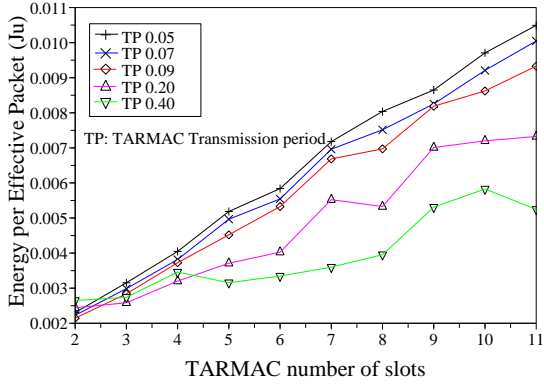}
\caption{TARMAC Energy Consumption Study}
\label{fig:busslots-enpkt}
\end{figure}

When the bus size increases, more collision happen and more packets
are dropped due to collisions. For those with longer periods, the
delivery ratio increases since each frame carries more packets.  After
some point, the collision ratio dominates as the physical capacity of
the channel is reached. Figure \ref{fig:busslots-occupacy} shows the
occupation ratio of TARMAC bus packets (effective packets / number of
slots). As can be expected, the longer the transmission period, the
higher the bus occupation ratio.
\begin{figure}[h]
\centering
\includegraphics[width=0.45\textwidth]{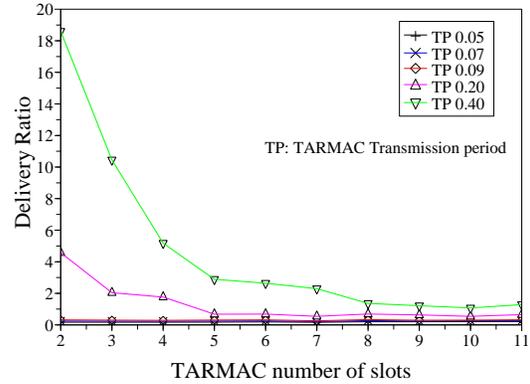}
\caption{TARMAC Average Delay Study}
\label{fig:busslots-delay}
\end{figure}
Figure \ref{fig:busslots-delay} shows the average delay of all
effective packets.  As the transmission period increases, the time to
get from source to destination increases. The smaller the buses are,
the longer is the time needed to reach the destination as packets wait
at each intermediate node.  Figure \ref{fig:busslots-enpkt} shows the
energy consumption per packet.  More energy is expended when the
period is reduced (sending more often) or when the frame size is
increased.

\subsection{TARMAC with different traffic patterns}

In this section, we study the performance of TARMAC under different
traffic patterns.  In the same grid scenario, we change the number of
data sources and their distribution in the networks, considering the
following 3 cases: all nodes as data sources, $\dfrac{1}{3}$($\sim
35$) nodes as data sources and a quarter of nodes in the same corner
are data sources. We also study two CBR send rates: 2 packets per
second and 1 packet per second.

\begin{figure}[h]
\centering
\includegraphics[width=0.45\textwidth]{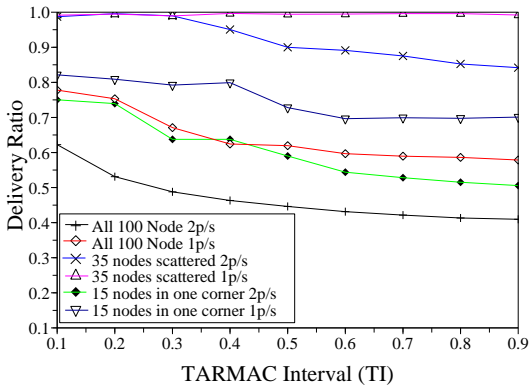}
\caption{TARMAC Delivery Ratio Study}
\label{fig:busptn-dr}
\end{figure}
\begin{figure}[h]
\centering
\includegraphics[width=0.45\textwidth]{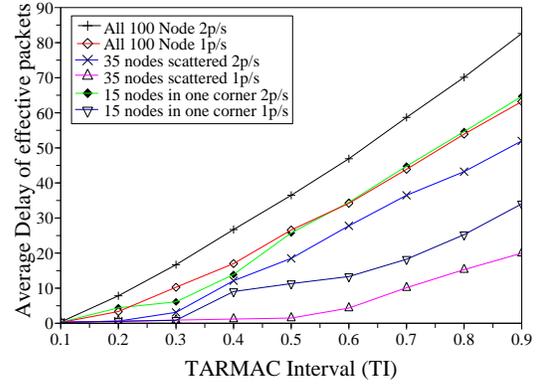}
\caption{TARMAC Delay Study}
\label{fig:busptn-delay}
\end{figure}
\begin{figure}[h]
\centering
\includegraphics[width=0.45\textwidth]{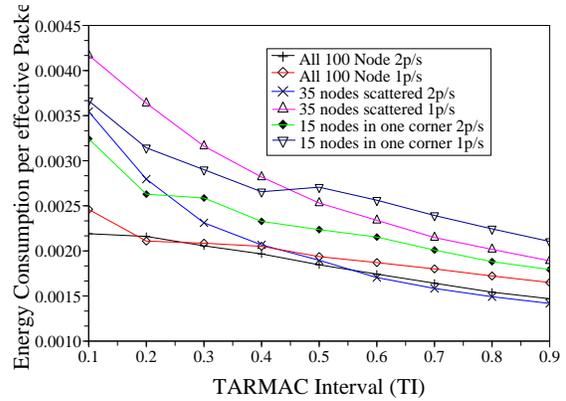}
\caption{TARMAC Average Energy Consumption Study}
\label{fig:busptn-enpkt}
\end{figure}

Figure \ref{fig:busptn-dr} shows the delivery ratios of TARMAC with
different traffic patterns.  As can be expected, the higher the data
rate, the lower the delivery ratio.  The capacity of the bus also
impacts the delivery ratio.  The delivery ratio of pattern with 35
nodes is higher than that of only 25 nodes. The reason is that those
25 nodes are in the same crowded area creating hotspotting along the
path to the basestation.  The average delay and energy consumption are
shown in figures \ref{fig:busptn-delay}, \ref{fig:busptn-enpkt} and
follow expectations.

\subsection{TARMAC vs Intrusion 1}

\begin{figure}[h]
\centering
\includegraphics[width=0.45\textwidth]{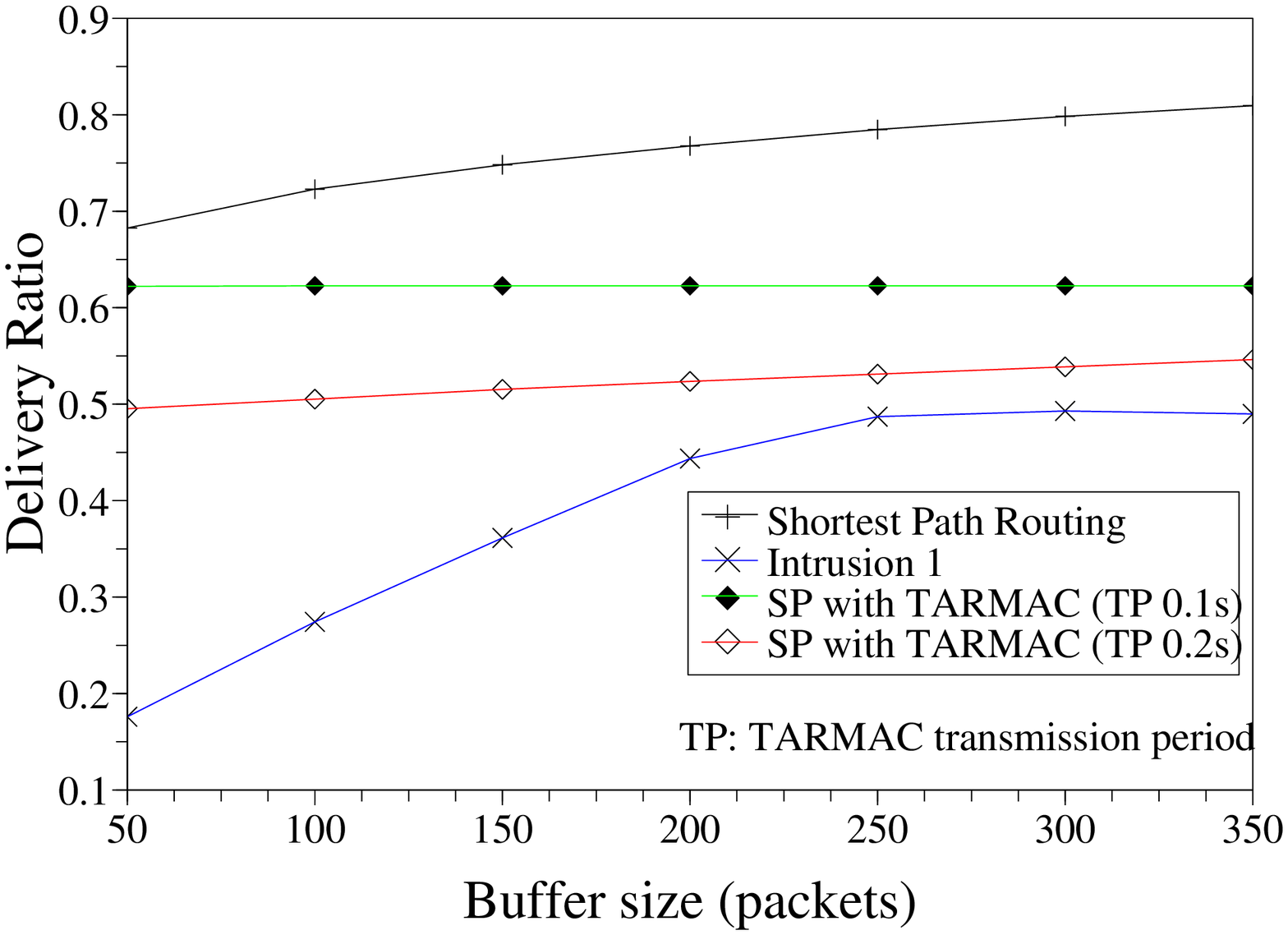}
\caption{Delivery Ratio}
\label{fig:bus-intrusion1_dr}
\end{figure}
\begin{figure}[h]
\centering
\includegraphics[width=0.45\textwidth]{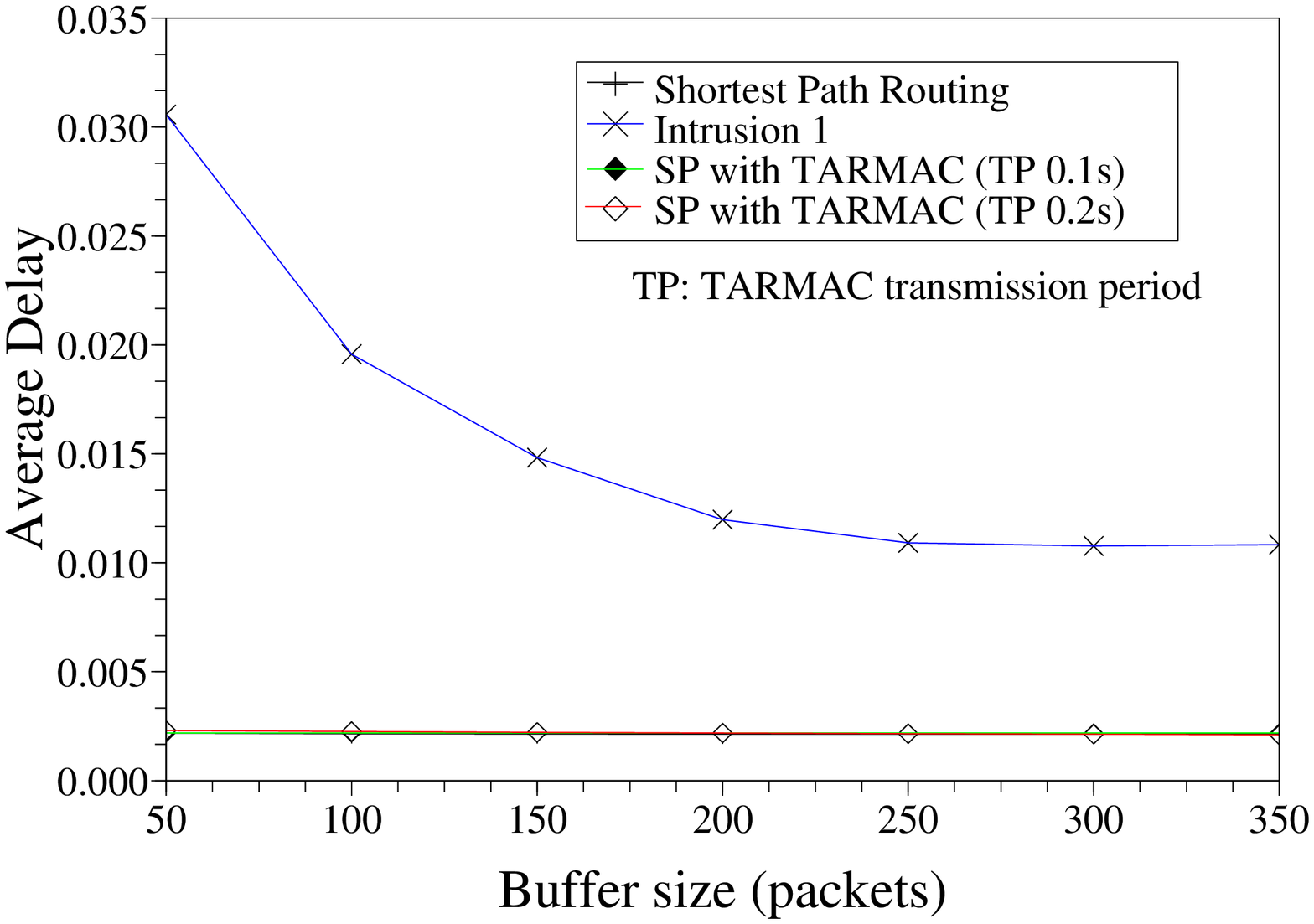}
\caption{Energy Consumption Per effective Packet}
\label{fig:bus-intrusion1_en}
\end{figure}
\begin{figure}[h]
\centering
\includegraphics[width=0.45\textwidth]{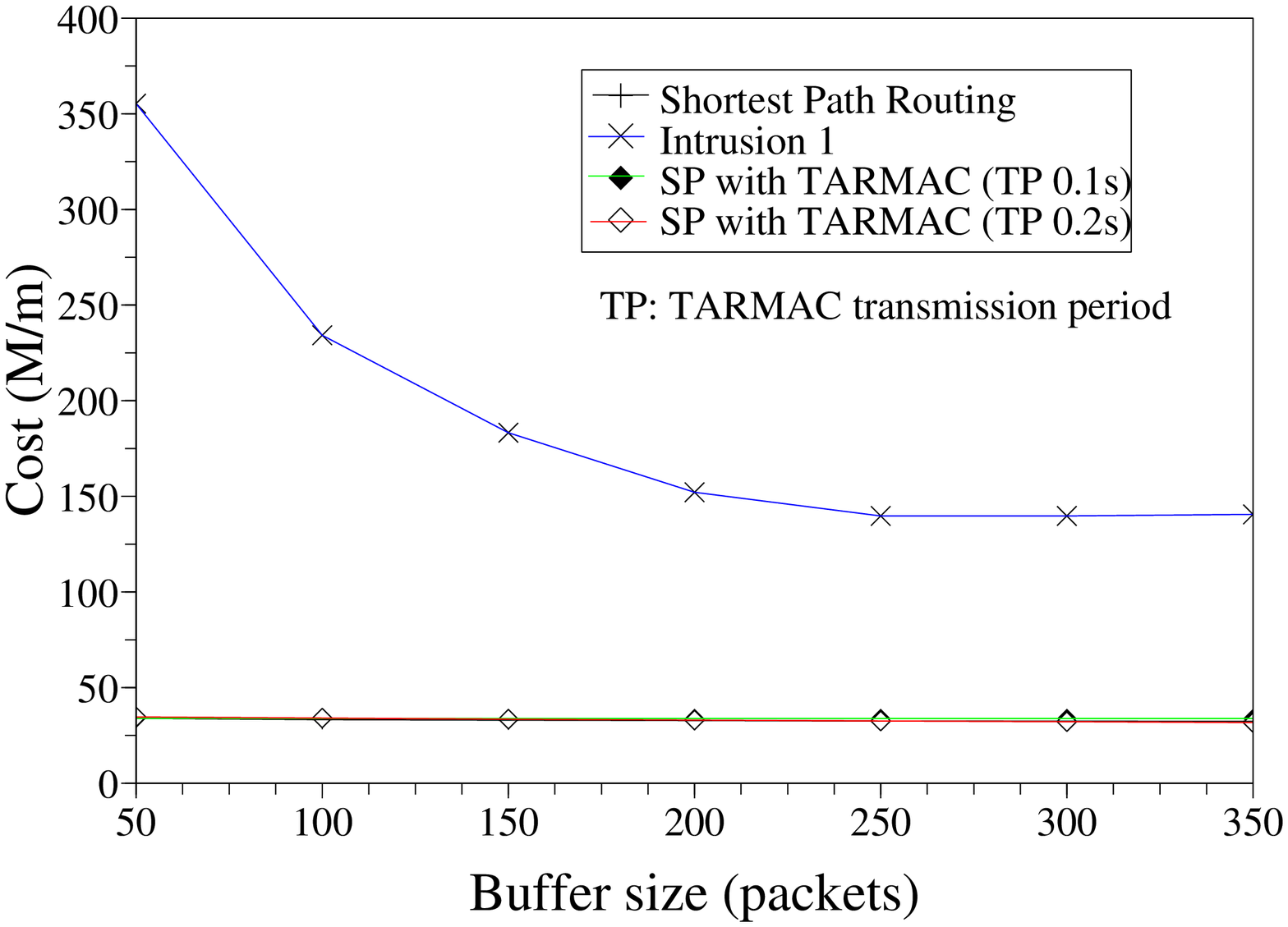}
\caption{Transmission Cost ($M/m$)}
\label{fig:bus-intrusion1_trans}
\end{figure}
\begin{figure}[h]
\centering
\includegraphics[width=0.45\textwidth]{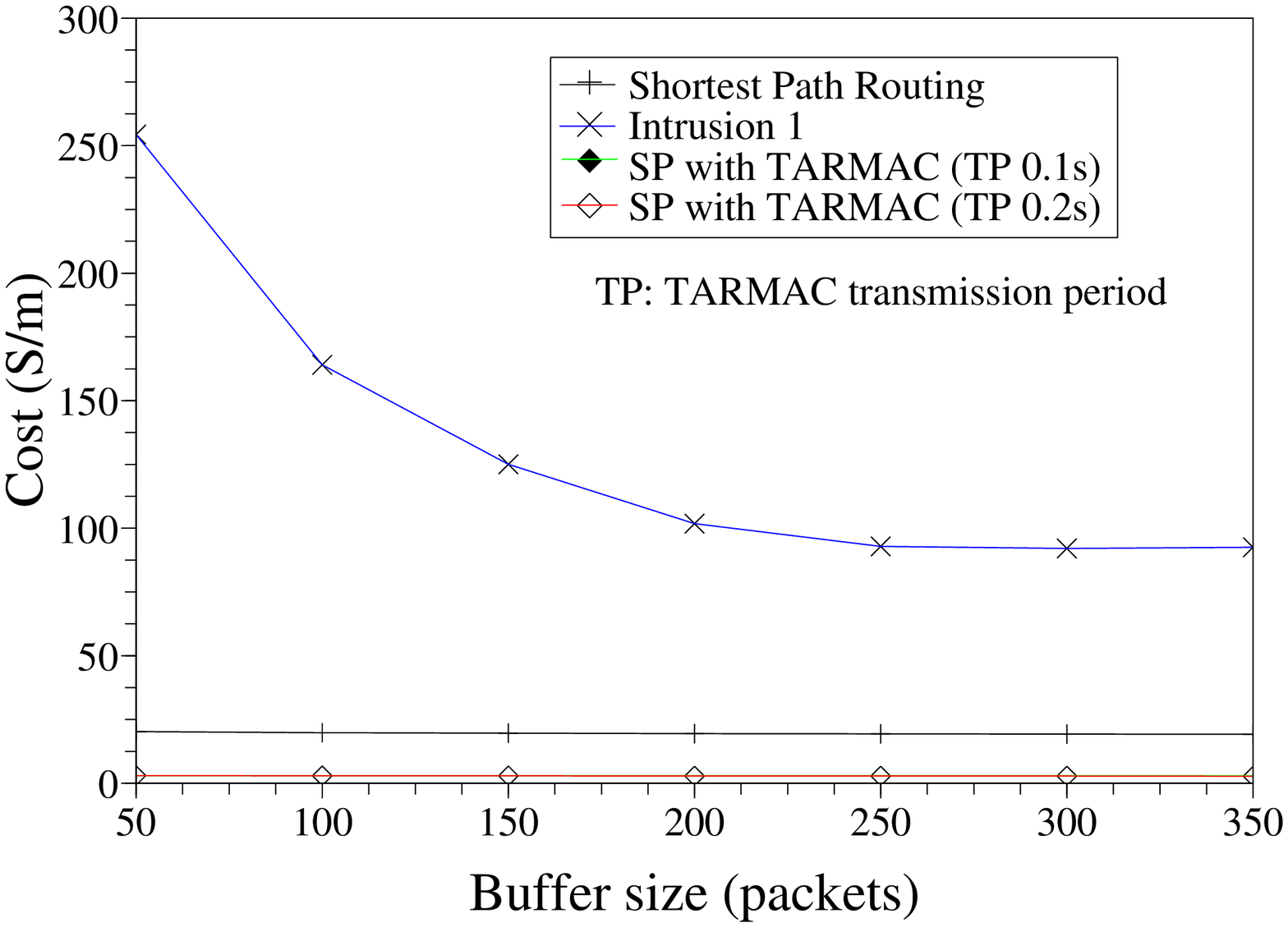}
\caption{Transmission Cost ($S/m$)}
\label{fig:bus-intrusion1_transize}
\end{figure}
\begin{figure}[h]
\centering
\includegraphics[width=0.45\textwidth]{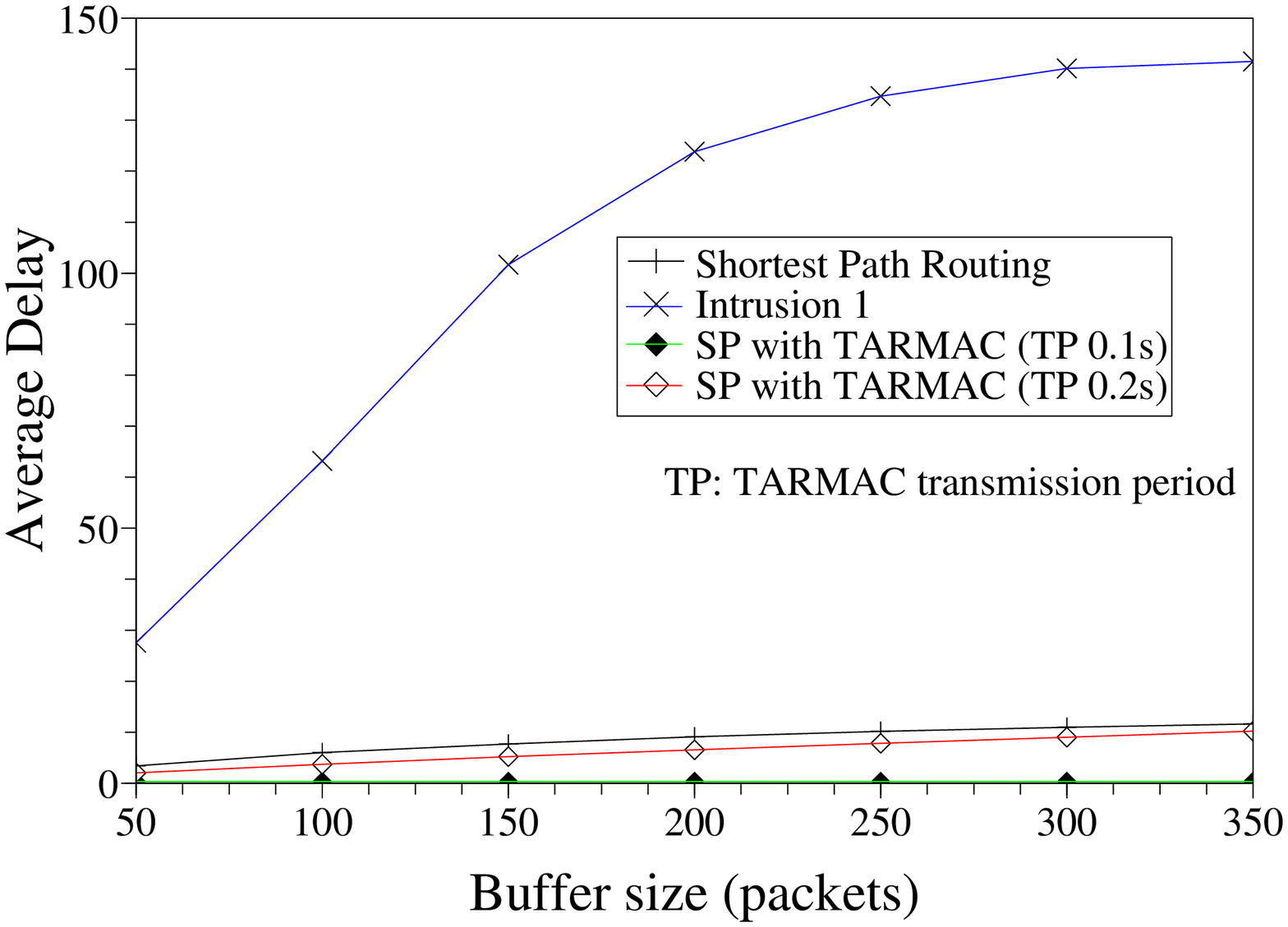}
\caption{Average End-to-End Transmission Delay}
\label{fig:bus-intrusion1_delay}
\end{figure}

In this study, we compare the performance of TARMAC and the Intrusion
1 solution~\cite{deng-intrusion}.   We select one node as a data sink in
the up-left corner of the simulation area.  Each node in the network
generates a CBR data traffic to the sink with a 2 packets per
second rate.

\rednote{Is the next commented stuff still true?}

Figure \ref{fig:bus-intrusion1_dr} shows the overall delay of TARMAC
and Intrusion 1, against that of Shortest Path routing which serves as
an upper bound on performance.  As we can see, the buffer size affects
the performance of intrusion 1 as well as bare SP. But TARMAC is not
affected apparently by the buffer size. The reason may be due to the
broadcast nature of TARMAC, and the unicast nature of intrusion 1 and
SP.  Since they use unicast, both intrusion 1
and SP may forward packet with best effort, holding packets for
possible retransmission until success. TARMAC only holds packet for
next available bus. So if the transmission period is low enough,
the required buffer size is not big.

Figure \ref{fig:bus-intrusion1_en} shows a somewhat surprising result
about the energy consumption per data packet received by the data
sink. The energy consumption per effective packet of TARMAC is nearly
the same as the ideal optimal solution. However, this result is
reasonable since each data packet is transmitted at each hop only
once, requiring nothing more than itself (in term of
transmission). Another reason leads to this result is the lower
delivery ratio. Some packets were dropped at the early stage of its
traveling from source to sink. We can reach this conclusion more
directly from Figure~\ref{fig:bus-intrusion1_trans} and
\ref{fig:bus-intrusion1_transize}.  Since unicast transmission used by
intrusion and SP requires control packets (RTS/CTS, ACK, etc.), the
number of packets transmitted physically is higher than that with
broadcasting: with TARMAC, all packets --routing, arp, data-- are
transmitted through broadcasting. $M/m$ refers to the total number of
packets transmitted physically (MAC layer packets) over the number of
effective packets (data packets received by sink).  $S/m$ indicates
the total size (in bytes) of packets transmitted physically over the
size of effective packets.

Figure \ref{fig:bus-intrusion1_delay} shows the average end-to-end
delay of all effective packets. The average delay of TARMAC is shorter
than it of SP may due to the lower delivery ratio (most packets with
longer delay are dropped before reaching the sink).

\subsection{TARMAC vs Intrusion 2}
\begin{figure}[h]
\centering
\includegraphics[width=0.45\textwidth]{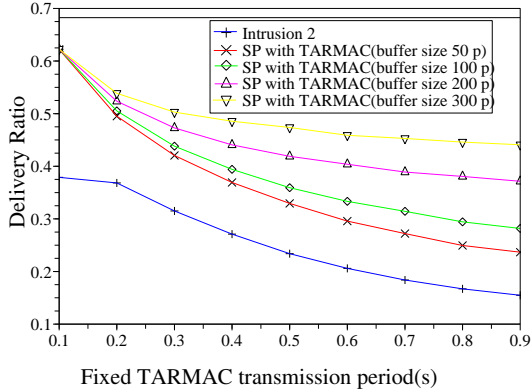}
\caption{Delivery Ratio}
\label{fig:bus-intrusion2_dr}
\end{figure}
\begin{figure}[h]
\centering
\includegraphics[width=0.45\textwidth]{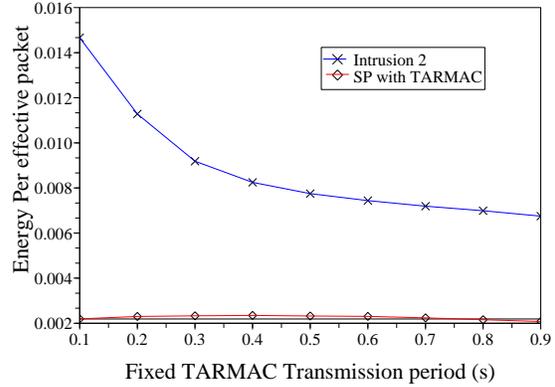}
\caption{Energy Consumption Per effective Packet}
\label{fig:bus-intrusion2_en}
\end{figure}

Intrusion 2~\cite{deng-antitrafficanalysis} does not buffer incoming
packets: only one packet is held for forwarding at each hop. Each node
keeps sending the held packet until it overhears the forwarding of the
corresponding packet at the next hop. Buffer size does not affect
the performance of TARMAC as well as was shown in the previous
section.%%  So we compare intrusion 2 with TARMAC with respect to the

Figure \ref{fig:bus-intrusion2_dr} shows the delivery ratio of TARMAC
and intrusion 2 with respect to the fixed transmission rate. The black
straight line stands the delivery ratio of ideal optimal solution
(SP). As the fixed transmission rate decreases, the delivery ratio
decreases as well since the capacity of the bus decreases.  As we can
see if the buffer size is double of TARMAC, the performance is
considerably better. Intrusion 2 can not benefit the increasing of the
buffer size due to its design is just holding one packet. Figure
\ref{fig:bus-intrusion2_en} shows the energy consumption per effective
packet.  The situation is the same as we discussed before. Intrusion 2
tries to retransmit a data packet many times at each hop through the
forwarding path. Meanwhile, no matter what the buffer size is, TARMAC
just forwards each packet once at each hop.

\begin{figure}[h]
\centering
\includegraphics[width=0.45\textwidth]{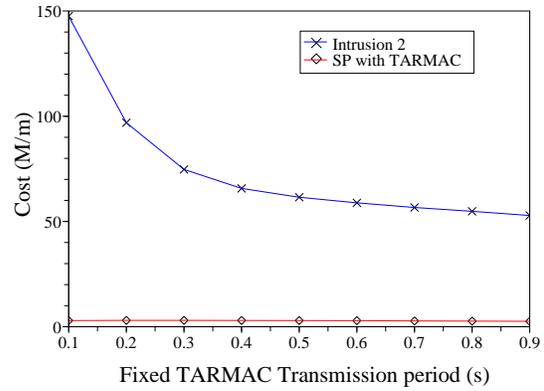}
\caption{Transmission Cost ($M/m$)}
\label{fig:bus-intrusion2_trans}
\end{figure}
\begin{figure}[h]
\centering
\includegraphics[width=0.45\textwidth]{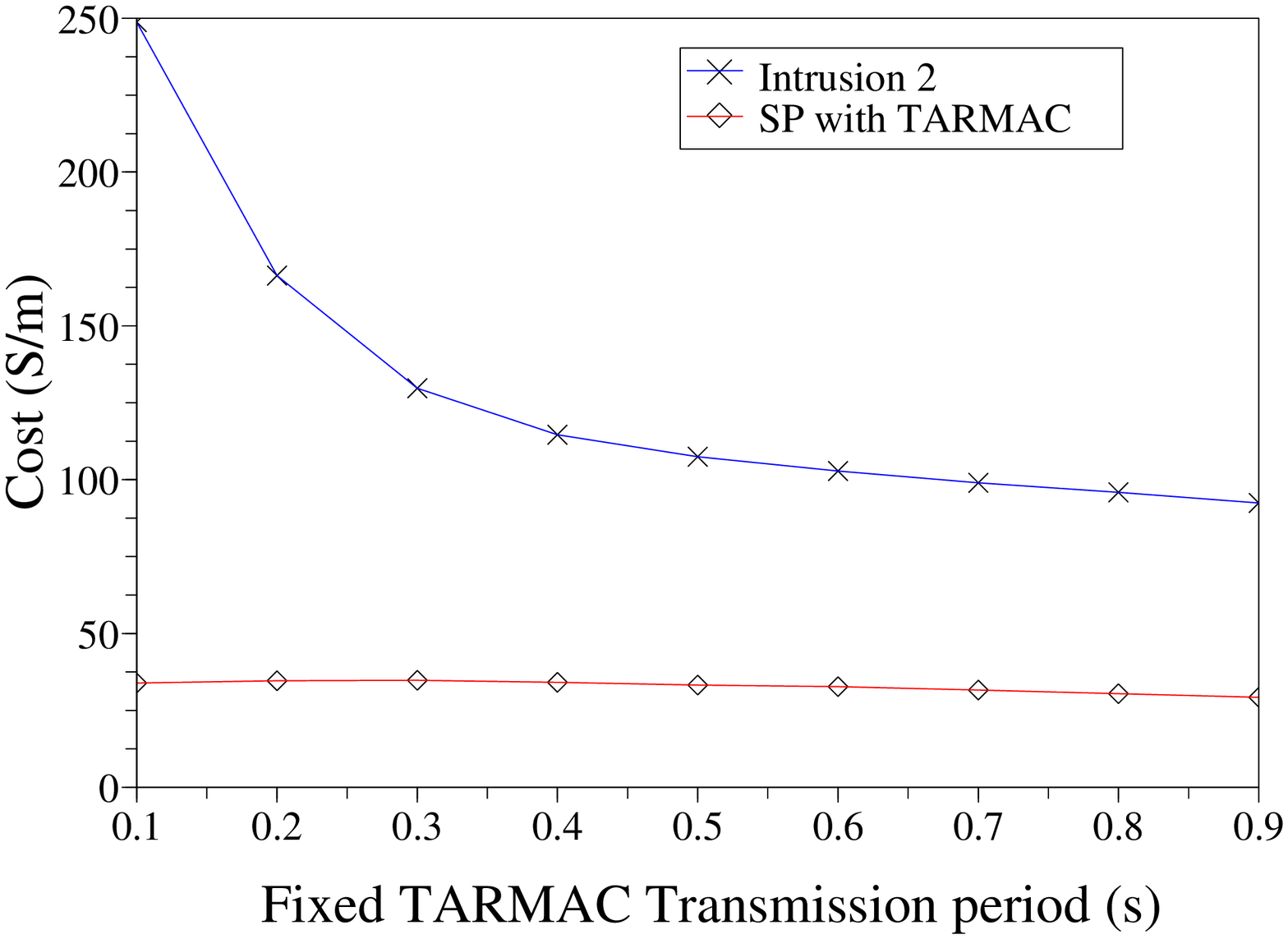}
\caption{Transmission Cost ($S/m$)}
\label{fig:bus-intrusion2_transize}
\end{figure}

\begin{figure}[h]
\centering
\includegraphics[width=0.45\textwidth]{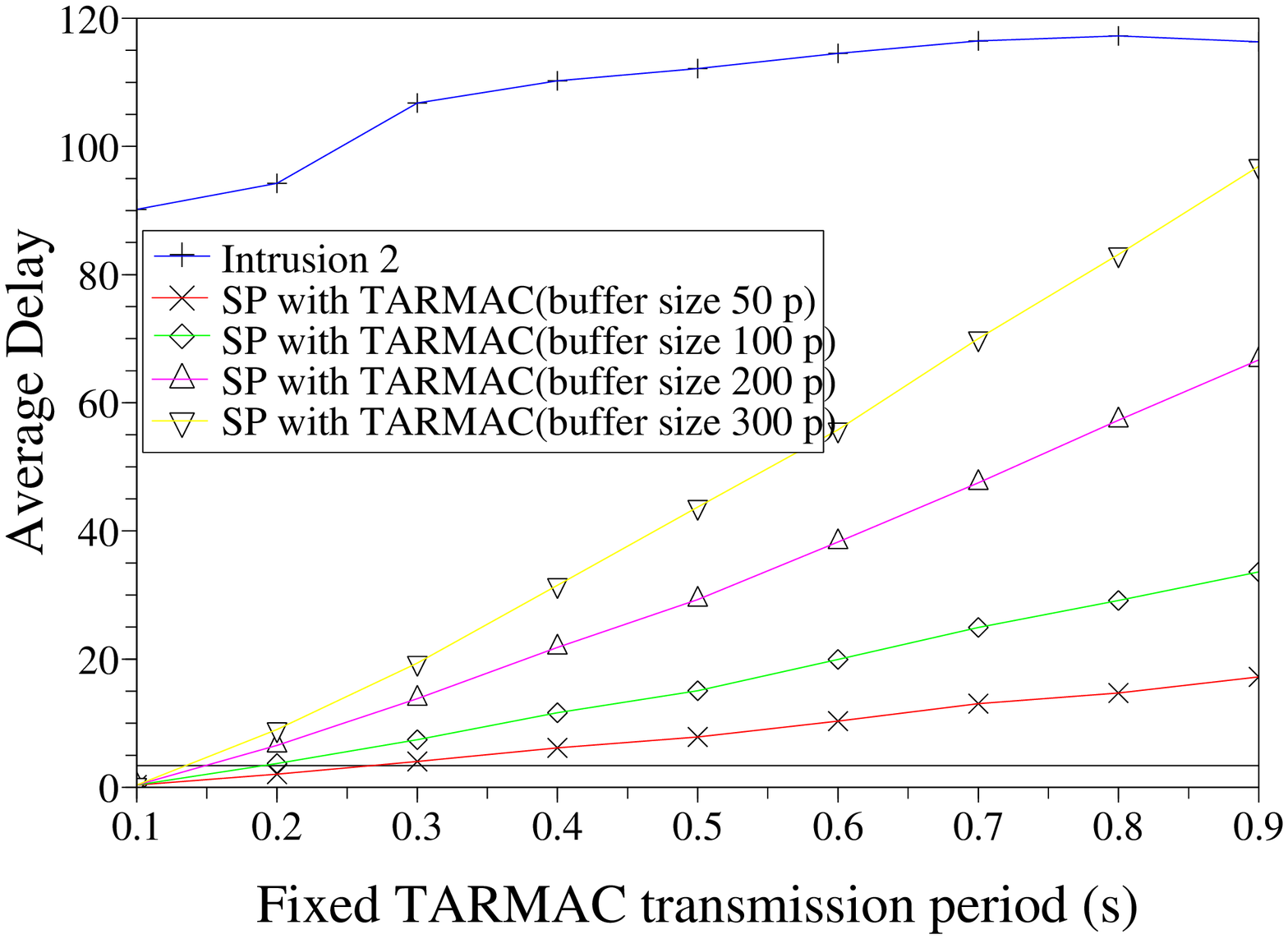}
\caption{Average End-to-End Transmission Delay}
\label{fig:bus-intrusion2_delay}
\end{figure}

Figure \ref{fig:bus-intrusion2_delay} shows the average end-to-end
delay of all effective packets. The straight line at bottom is that of
the ideal solution (SP).  At first, the average delay of TARMAC with
all buffer sizes are shorter than that of SP, perhaps due to the lower
delivery ratio: most data packets with longer delay were dropped. As
the buffer sizes increase, the average delay of TARMAC increases.

\subsection{Evaluation of Multipath Routing and Adaption of TARMAC frame frequency}
The routing layer is orthogonal to TARMAC, however, if it makes uses TARMAC it
can greatly increase its performance. Let us assume that the routing layer just
implements, say, shortest path routing. Now, if there is an event at a certain
location and all the nodes at that location start sending data suddenly, the
capacity of the network will be reached and the queues would fill up causing packets
to be dropped. In addition to this, the empty TARMAC frames from nodes adjacent to
the data flow would interfere with the data flow causing further drops. However,
if the routing layer took advantage of the presence of TARMAC, it could employ multi-path
routing. This would mean that the frames that were originally empty would now hold data packets
and portions of the data would follow different paths to get to the destination. This would
increase the delivery ratio, reduce the load on the network capacity and decrease the overhead
as now the frames are carrying packets rather than being empty.
\begin{figure}[h]
\centering
\includegraphics[width=0.45\textwidth]{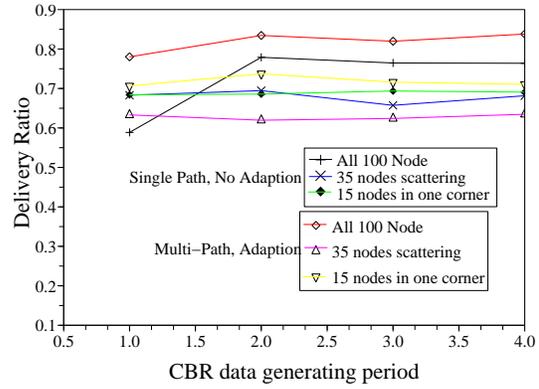}
\caption{Multipath Routing and Adaption gain}
\label{fig:mp-adapt-dr}
\end{figure}
The figure \ref{fig:mp-adapt-dr} shows the effects of multipath routing and the
adaption of frame frequency. The adaption and multipath routing helps more packets passing
through the network if density of data sources are high. The does not help to much (or
even hurt) those with lower data sources density, since more collision may happen.

\section{Concluding Remarks} \label{conclude}

In this paper, we proposed a novel traffic-analysis resitant MAC
protocol that replaces typical data-driven protocols with a fixed
transmission schedule.  TARMAC can interoperate with any routing protocol (although
one that is aware of its characteristics can best utilize it).  Further, while
we evaluated TARMAC with a contention based MAC, it can interoperate with other link
layer protocols, including TDMA based ones, or be extended to support reliability.

We evaluated the characteristics of this scheme, compared to 
existing approaches and showed that it offers significant performance
advantages.  Further, we discussed improvements to the basic TARMAC to further
improve its efficiency without compromising its resilience to traffic analysis.

\bibliographystyle{IEEEtran}
\bibliography{references}

\begin{thebibliography}{1}
\providecommand{\url}[1]{#1}
\csname url@rmstyle\endcsname
\providecommand{\newblock}{\relax}
\providecommand{\bibinfo}[2]{#2}
\providecommand\BIBentrySTDinterwordspacing{\spaceskip=0pt\relax}
\providecommand\BIBentryALTinterwordstretchfactor{4}
\providecommand\BIBentryALTinterwordspacing{\spaceskip=\fontdimen2\font plus
\BIBentryALTinterwordstretchfactor\fontdimen3\font minus
  \fontdimen4\font\relax}
\providecommand\BIBforeignlanguage[2]{{%
\expandafter\ifx\csname l@#1\endcsname\relax
\typeout{** WARNING: IEEEtran.bst: No hyphenation pattern has been}%
\typeout{** loaded for the language `#1'. Using the pattern for}%
\typeout{** the default language instead.}%
\else
\language=\csname l@#1\endcsname
\fi
#2}}

\bibitem{deng-intrusion}
J.~Deng, R.~Han, and S.~Mishra, ``Intrusion tolerance and anti-traffic analysis
  strategies for wireless sensor networks,'' in \emph{DSN '04: Proceedings of
  the 2004 International Conference on Dependable Systems and Networks
  (DSN'04)}.\hskip 1em plus 0.5em minus 0.4em\relax Washington, DC, USA: IEEE
  Computer Society, 2004, p. 637.

\bibitem{mathewson-practical}
\BIBentryALTinterwordspacing
N.~Mathewson and R.~Dingledine, ``Practical traffic analysis: Extending and
  resisting statistical disclosure.'' [Online]. Available:
  \url{citeseer.ist.psu.edu/678780.html}
\BIBentrySTDinterwordspacing

\bibitem{newman-metrics}
\BIBentryALTinterwordspacing
R.~E. Newman, I.~S. Moskowitz, P.~Syverson, and A.~Serjantov, ``Metrics for
  traffic analysis prevention.'' [Online]. Available:
  \url{citeseer.ist.psu.edu/653919.html}
\BIBentrySTDinterwordspacing

\bibitem{syverson97anonymous}
\BIBentryALTinterwordspacing
P.~F. Syverson, D.~M. Goldschlag, and M.~G. Reed, ``Anonymous connections and
  onion routing,'' in \emph{{IEEE} Symposium on Security and Privacy}, Oakland,
  California, 4--7 1997, pp. 44--54. [Online]. Available:
  \url{citeseer.ist.psu.edu/syverson97anonymous.html}
\BIBentrySTDinterwordspacing

\bibitem{deng-antitrafficanalysis}
J.~Deng, R.~Han, and S.~Mishra, ``Countermeasures againts traffic analysis
  attacks in wireless sensor networks,'' in \emph{SecureComm '05: CerateNet
  Conference on Security and Privacy in Communication Networks (SecureComm
  2005)}.\hskip 1em plus 0.5em minus 0.4em\relax IEEE Computer Society, 2005,
  pp. 113--124.

\bibitem{lee-01}
S.~Lee and M.~Gerla, ``Split multipath routing with maximally disjoint paths in
  ad hoc networks,'' in \emph{Proceedings of the IEEE International Conference
  on Communication (ICC)}, 2001, pp. 3201--3205.

\bibitem{xu-01}
Y.~Xu, J.~Heidemann, and D.~Estrin, ``Geography-informed energy conservation
  for ad hoc routing,'' in \emph{MobiCom '01: Proceedings of the 7th annual
  international conference on Mobile computing and networking}.\hskip 1em plus
  0.5em minus 0.4em\relax New York, NY, USA: ACM Press, 2001, pp. 70--84.

\bibitem{ns2}
\BIBentryALTinterwordspacing
``Network simulator - ns2.'' [Online]. Available:
  \url{http://www.isi.edu/nsnam/ns/}
\BIBentrySTDinterwordspacing

\end{thebibliography}
\end{document}